# Excitation Wavelength Dependent Reversible Photoluminescence Peak in Iodide Perovskites

Wayesh Qarony[1], Mohammad K. Hossain[3], Mohammad I. Hossain[1], Sainan Ma[1], Longhui Zeng[1], Kin Man Yu[3], Dietmar Knipp[4], Alberto Salleo[4], Huarui Sun[2], Cho Tung Yip[*, 2], Yuen Hong Tsang[*, 1]

1.) Department of Applied Physics, The Hong Kong Polytechnic University, Hung Hom, Kowloon, Hong Kong

2.) School of Science, Harbin Institute of Technology, Shenzhen Graduate School, Shenzhen, Guangdong 58055, China

3.) Department of Physics, City University of Hong Kong, Kowloon, Hong Kong

4.) Geballe Laboratory for Advanced Materials, Department of Materials Science and Engineering, Stanford University, Stanford, CA 94305, USA

*Corresponding authors: E-mail: yuen.tsang@polyu.edu.hk (Y. H. Tsang), h0260416@hit.edu.cn (C. T. Yip)

**ABSTRACT**

Halide perovskites have indisputably exceptional optical and electronic properties, which are attractive for next-generation optoelectronic device technologies. We report on a reversible photoluminescence (PL) peak in iodide-based organic-inorganic lead halide perovskite materials under a two-photon absorption (TPA) process, while tuning the excitation wavelength. This phenomenon occurs when the incoming femtosecond (fs) laser photon energy is higher than a threshold energy. Intriguingly, this phenomenon also occurs in other kinds of iodide perovskite materials. Moreover, two more shorter wavelength peaks exhibit and become prominent when the excitation photon energy is being tuned in the high energy wavelength spectrum, while laser power is remained constant. However, the spectral PL energy window between the original material peak and the first high energy peak can vary based on the optoelectronic properties of the prepared films. The same phenomenon of reversible PL peak is also observed in various iodide based organic-inorganic halides as well as all-inorganic perovskite single crystals and polycrystals. We attribute to the reversible PL peak to the photoinduced structural deformation and the associated change in optical bandgap of iodide perovskites under the femtosecond laser excitation. Our findings will introduce a new degree of freedom in future research as well as adding new functionalities to optoelectronic applications in these emerging perovskite materials.

Recently, perovskites have drawn substantial research interest in optical gain media of light emitting devices [1-15] owing to their excellent optoelectronic characteristics [1,16-17], after making an unprecedented success story in solar cells [28–33]. Researchers in the past decades have studied the lasing properties in different types of organic-inorganic and all-inorganic lead halide perovskites, such as perovskite films [6,7], nanowires [2,8–10], micro/nanoplatelets [11–13,23], and quantum dots [6,14,15,26]. Perovskites are basically a group of materials with a general formula of $ABX_3$, where A is an organic/inorganic cation (organic: $CH_3NH_3^+$ ($MA^+$) or $NH_2CH_3NH_2^+$($FA^+$), inorganic: Cs), B is a divalent anion ($Pb^+$ or $Sn^+$), and X is a monovalent halide anion ($I^-$, $Cl^-$, or $Br^-$) [3]. The halide atoms are located in eight-sided (octahedra) structures, while the divalent anions of lead or tin atoms sit inside the octahedra and the organic/inorganic anions take the position between octahedra. So far, the outstanding perovskite optoelectronic devices are mainly observed in iodide-based hybrid halide perovskites. However, the presence of the iodine element in the halide perovskites creates degradation and instability problems under illumination [34–36]. Nevertheless, the light exposure on the iodide perovskites might also have some positive impacts. For example, Wu et al. reported that the iodine atoms are rotated around the lead atom under the photoexcitation, leading to switch of each octahedron structure from regular shape to a distorted shape (see Fig. 4(a)) [37]. This fs laser-induced distorted structure could facilitate an enhanced migration of charge carriers through defects and prevent them from being trapped, allowing to obtain higher energy conversion efficiency in perovskite solar cells [37]. Moreover, laser-induced surface relaxation/derelaxation pronouncing a reversible spectral evolution and PL redshift/blueshift as a function of continuous wave (CW) laser annealing is recently demonstrated in iodide-based 2D Ruddlesden–Popper perovskites (RPP) [38]. However, the fs optical pumped laser maintains a relatively much broader optical spectral bandwidth with higher laser intensity in comparison with the CW laser, allowing to demonstrate a variety of novel and exciting effects mainly due to the laser-induced ultrafast structural changes. Such photoinduced non-thermal structural transformation is demonstrated in two-dimensional (2D) layered materials, where diamond/graphite can be changed to graphite/graphene under fs laser pulse excitations [35]. Besides, an anomalous blueshift with a reversible structural change dynamic is observed in $[Ge_2Te_2/Sb_2Te_3]_{20}$ materials under a strong double-pulse excitation along with employing the fs laser source [39]. Furthermore, an analysis based on DFT calculation of the structural deformed $CH_3NH_3PbI_3$ perovskites' electronic properties confirms that the iodine elements in the network of Pb-I-Pb are chemically unstable due to the difference in bond lengths of one Pb-I to another Pb-I framework [40]. This may cause to the possible structural deformation under ultrafast photoexcitation. Despite obtaining such findings, some of the fundamental working mechanisms and phenomena responsible for the exciting and unique performances, particularly, in iodide perovskite devices are not completely understood [10,18,41,42]. This intrigues us to further study on the iodide-based perovskites under the photoexcitation using fs laser.

In this work, two-photon photoluminescence (TPL) spectroscopy on different kinds of perovskite materials is conducted to study their two-photon absorption (TPA) processes. The investigation is carried out for both organic-inorganic and all-inorganic polycrystalline thin films, and single crystal nanostructures of random shapes (rectangular-like, square-like, triangular, etc.) including low-dimensional lyered crystals of nanowires. We report on a phenomenon of excitation wavelength-dependent fully reversible PL peak in iodide-based hybrid perovskite materials under a TPA absorption process, where a fs laser is utilized as a photoexcitation source.

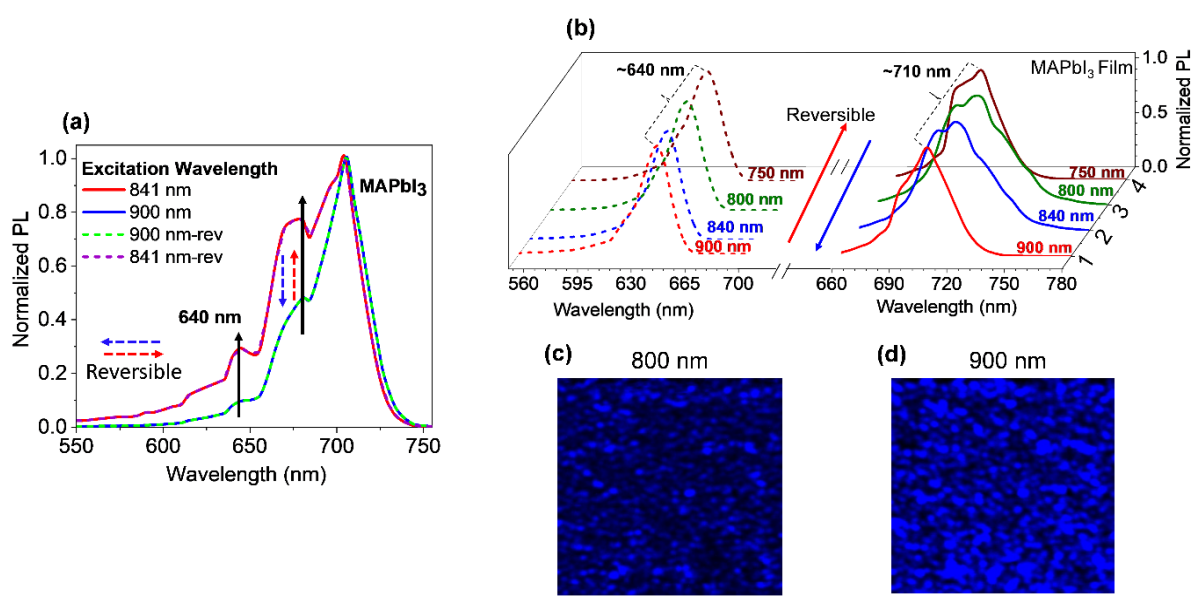

*Figure 1| Reversible photoluminescence peak in iodide-based perovskites: **a**, An experimentally realized integrated PL (normalized) spectral evolution of $MAPbI_3$ perovskite for the tunable TPA pumped lasing excitation wavelengths of 900 nm and 841 nm. The sample exhibits a PL peak about 710 nm for the excitation of 900 nm. Two additional peaks (~645 nm and ~680 nm) exhibit and become more prominent with the excitation of shorter wavelength, while the laser intensity/power was kept constant. When the excitation changes from one wavelength to another (e.g. 900 to 841 nm and vice versa) the corresponding PL peak also changes accordingly, which is stated here as reversible property of material. **b**, The*

*phenomenon with reversible behavior is comprehensively studied by varying the TPA excitation wavelength ranging from 900 nm to 750 nm, while a 680 nm of short-pass filter is utilized to clearly observe the effect on the image of ~640 nm PL peak (dotted PLs on the left) and to ensure that this emission is from perovskite rather than some artifacts. **(c,d)** Photographs of MAPbI$_3$ polycrystalline sample illuminated by (a) 800 nm and (b) 900 nm TPA laser beams. In particular, the photograph for the excitation of 800 nm nicely demonstrates the emission from the material.*

In the first part of this research, a systematic fs laser study is carried out on the most commonly-studied methylammonium lead iodide (CH$_3$NH$_3$PbI$_3$ / MAPbI$_3$) perovskite, where a TPA pumped lasing is used for the investigation. The MAPbI$_3$ perovskite thin film was synthesized by a two-step solution-processed method [40,41], where PbI$_2$ and CH$_3$NH$_3$I (MAI) were used as precursors. The scanning electron microscopic (SEM; JEOL, JEM-2100F) image revealed an average grain size over a micron scale (Fig. S1(a) inset). X-ray diffraction (XRD; Rigaku SmartLab) pattern (Fig. S1(a)) confirmed that the MAPbI$_3$ film is polycrystalline with a good crystalline quality. The PL measurements were performed on MAPbI$_3$ film/FTO configuration while tuning the fs laser wavelength from 700 nm to 900 nm at room temperature. The PL and optical image measurements were conducted by a laser scanning multiphoton (MP) confocal microscope system (TCS SP8, Leica) combined with a Ti: sapphire fs laser (Mai Tai HP, Spectra-Physics) excitation. The laser pulse repetition rate and the time duration were 80 MHz and 500 fs, respectively, while the spectrometer was 5 nm. An integrated hybrid detector of the laser scanning confocal microscope system was utilized to detect the nonlinear emission signal. PL spectra emitted from the MAPbI$_3$ film is also included in the supplementary Fig. S1(b), while the pumping laser power is varied from 0.1 mW to 14.0 mW. The inset Fig. S1(b) demonstrates the evolution from a spontaneous emission to lasing behavior over the complete spectral range as a function of pumping power.

Figure 1(a) displays the TPA excitation wavelength-dependent spectral evolution of as prepared MAPbI$_3$ perovskite film, while the excitation wavelength is tuned within a particular spectral range. The PL peak pronouncing at ~710 nm for the TPA excitation wavelength of 900 nm corresponds to the optical bandgap of the prepared sample. Interestingly, when the excitation changes from longer wavelength to relatively shorter wavelength (e.g. 900 to 841 nm), two additional peaks exhibit locating at ~640 nm and ~680 nm as shown in Fig. 1(a). Those two peaks become more prominent with the decrease of excitation wavelength or increase of excitation energy, while the excitation laser intensity or power is remained constant. In fact, the room temperature power-dependent PL spectra of CH$_3$NH$_3$PbI$_3$/ CsPbBr$_{0.15}$I$_{0.85}$ films at 900 nm of TPA lasing wavelength demomstrates that the laser power/intensity does not have such infleuce on the PL spectra as provided in the supplementary file (Fig. S1(b) & S1(d)). More interestingly, this anomalous PL peak evolution process is completely reversible between any two distinct excitation wavelength when the TPA excitation wavelength is tuned around them e.g. the PL emission peak without additional prominent peaks for the two-photon excitation of 900 nm laser can be fully resolved back after the two-photon excitation and emission of 841 nm and with prominent PL peaks and vice versa as shown in Fig. 1(a).This phenomenon with reversible behavior is further studied by varying the TPA excitation wavelength ranging from 900 nm to 750 nm as displayed in Fig. 1(b). In this case, a 680 nm of short-pass filter is utilized to clearly observe the exhibition of ~640 nm PL peak with better images, while excitation wavlengths are varied. The use of filter also makes sure that the emission is actually originated from the perovskites rather than any artifact, which is nicely demonstrated from the photograph of perovskite film under the excitation of 800 nm as shown in Fig. 1(c). Hence the 680 nm optical short-pass filter is used for the following complete study of this manuscript. The normalized PL peak exhibited at ~640 nm with the variation of TPA excitation wavelength (900 nm to 750 nm) is depicted in Fig. 1(b) (left with dotted peaks). The investigation was mainly conducted here focusing on the first high energy peak (640 nm), since this PL peak can be pronounced in other kinds of perovskite materials as well, which is described in the later part of this manuscript. In this case, when the MAPbI$_3$ films are illuminated by the irradiation wavelengths, the two peaks exhibited at 710 nm and ~640 nm are separated with a spectral energy window of $\Delta E_{PL}$~ 0.2 eV (($\Delta \lambda_{PL}$=70 nm). The exhibition of these two peaks with a certain spectral energy window is also reversible with the excitation of wavelength. In fact, the spectral PL energy window between the original material peak and the first high energy peak can vary based on the optoelectronic properties of the prepared films, which will be again discussed in the following. This phenomenon always occurs when the incoming fs laser photon energy is higher than a fixed threshold energy.

It is assumed that energy bands are relaxed/derelaxed under certain excitation energy and hence the lowest vibrational level of conduction bands and/or highest vibrational level of valance bands are splitted and/or shifted when the photoexcitation energy is tuned around a threshold level. As the perovskite film is excited by a photon energy of TPA excitation wavelength, the PL center as usually relaxes from the maximum level of conduction energy band to the lowest vibrational level of the band and emits excess energy to the surroundings by returning to the ground energy. The further photoexcitation with relatively higher energy or shorter TPA excitation wavelength may create structural changes in the material [35,37,39]. Consequently, the lowest level of the vibrational conduction bands (CB) or highest level of the valance bands (VB) can be splitted and then energy states can be shifted up (CB) or down (VB), resulting in photoemission with relatively higher energy along with the original emission. Intriguingly, the previous state can be reversed back if the sample is subjected to the relatively lower energy excitation wavelength (e.g. 900 nm). In other words, a fully reversible energy emission process can be entirely controlled by the incident two-photon energy from the fs laser.

*Table 1: Different investigated perovskite materials with the existence/non-existence of iodine element, preparation methods, microstructures, and the energy window between two prominent peaks .*

| Perovskites | Preparation method | Microstructure | PL energy window ($\Delta E_{PL}$) |
|---|---|---|---|
| $MAPbI_3$ | solution-processed | polycrystal (film) | Yes, 710->640 nm (~0.19 eV) |
| $FA_xMA_{1-x}PbI_3$ | solution-processed | polycrystal (film) | Yes, 707->645 nm (~0.17 eV) |
| $CsPbBr_3$ | solution-processed | polycrystal(film) | No, **530 nm** (~0.0 eV) |
| $FASnI_3$ | solution-processed | polycrystal(film) | Yes, 705->640 nm (~0.18 eV) |
| $MAPbI_3$ | solution-processed | single-crystal nanowire | Yes, 715->645 nm (~0.19 eV) |
| $CsPbBr_{0.4}Cl_{0.6}$ | CVD | single-crystal microplatelets | No, **475 nm** (~0.0 eV) |
| $CsPbBr_{0.15}I_{0.85}$ | CVD | single-crystal microplatelets | Yes, 665->640 nm (~0.07 eV) |
| $CsPbBr_{0.08}I_{0.92}$ | CVD | single-crystal microplatelets | Yes, 700->645 nm (~0.17 eV) |

With the aim of investigating the physical origin of the phenomenon, different solution-processed and chemical vapor deposited (CVD) organic-inorganic and all-inorganic perovskites (polycrystalline films and single crystal microplatelets) are synthesized and characterized under the similar condition using fs laser system. The comprehensive results of those investigated perovskites are summarized in Fig. 2 and Table 1, where the 680 nm short-pass filter was utilized only for the case of 800 nm excitations. The photographs of CVD-processed single crystal perovskites illuminated by 900 nm laser beam are displayed in Fig. 2(e), whereas photographs of all other investigated films are included in the supplementary document (see Fig. S2). Similar to the solution-processed MAPbI$_3$ film, the spectral evolution of reversible PL peak and the exhibition of reversible two prominent peaks (original and the 1$^{st}$ high energy) are observed in all the iodide-based perovskites regardless of lead- or lead-free perovskites, preparation methods (solution-processed/CVD), composition, and microstructures (polycrystal/single crystal) of the prepared samples as presented in Fig. 2. However, the extent of the energy window between two prominent peaks varies based on the optical properties of the prepared perovskite samples. The spectral energy window of FA$_x$MA$_{1-x}$PbI$_3$(film), FASnI$_3$ (film), MAPbI$_3$(single-crystal NW), CsPbBr$_{0.15}$I$_{0.85}$ (single-crystal microplatelets), and CsPbBr$_{0.08}$I$_{0.92}$ (single-crystal microplatelets) perovskites are approx. 0.17 eV, 0.18 eV, 0.19 eV, 0.07 eV, and 0.17 eV, respectively as depicted in Fig. 2 (a, b, c, d, e) and tabulated in Table 1.

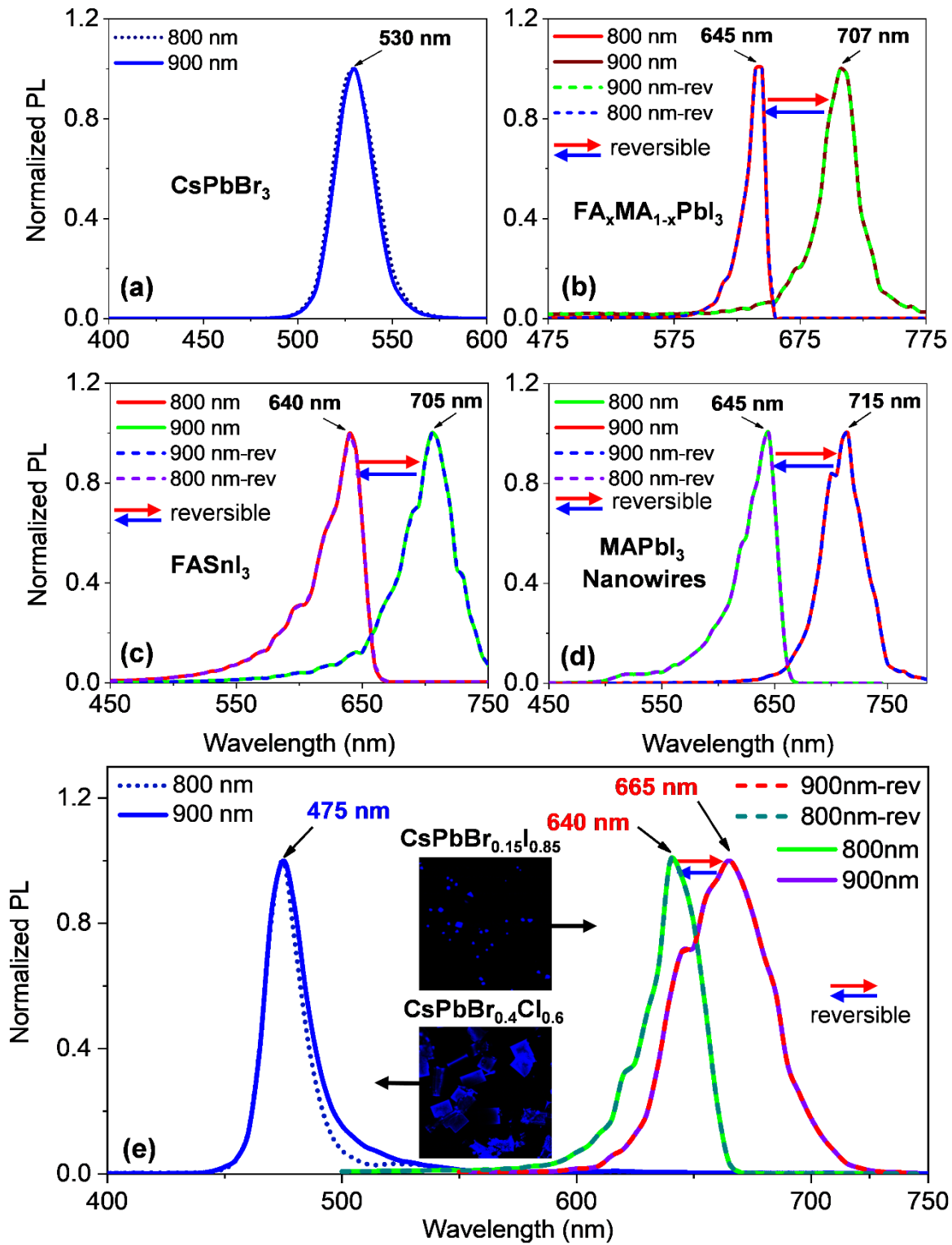

*Figure 2| Comprehensive demonstration of reversible PL with two different emission energy peaks in iodide perovskites. (a-d),* Different organic-inorganic (MAPbI$_3$, FASnI$_3$, and FA$_x$MA$_{1-x}$PbI$_3$) and all-inorganic (CsPbBr$_3$) lead/tin halide solution-processed perovskites are investigated for two different excitaton wavelengths of 800 nm and 900 nm, where a 680 nm

*short-pass filter was utilized for the case of 800 nm. (**a**) all-inorganic CsPbBr$_3$ polycrystal perovskite does not exhibit excitation wavelength-dependent reversible PL characteristics as well as reversible two PL peaks, whereas (**b**) FA$_x$MA$_{1-x}$PbI$_3$ polycrystal, (**c**) FASnI$_3$ polycrystal, and (**d**) MAPbI$_3$ single crystal nanowire perovskites exhibit the similar phenomenon to MAPbI$_3$. **e**, Further investigation in all-inorganic single crystal perovskite alloys with/without the halide element of iodine (I) pronounce the same. The insets show the photographs of CsPbBr$_{0.15}$Cl$_{0.85}$ and CsPbBr$_{0.4}$Cl$_{0.6}$ (top -> bottom) microplatelets single crystal illuminated by a 900 nm laser beam.* **The above comprehensive study (a-e) concludes that the excitation wavelength-dependent reversible PL and reversible two prominent peaks phenomenon are pronounced only for the perovskites with the element of iodine.**

Nevertheless, unlike iodide perovskites demonstrated above, all-inorganic CsPbBr$_3$ (film, solution-processed) and CsPbBr$_{0.4}$Cl$_{0.6}$ (single-crystal alloy microplatelets, CVD) perovskites, without iodine anion, show different photoluminescence emission behavior, revealing only one PL peak located at 530 nm and 475 nm, respectively as shown in Fig. 2 (a, e) and Table 1. By tuning the excitation photon energy around the threshold energy, they do not show any further emission peak, indicating the absence of additional energy state. Hence, it is reasonable to infer that the phenomenon of reversible PL and reversible two prominent peaks with a certain energy gap under fs laser excitations only occurs in iodide-based hybrid perovskites, irrespective if the perovskite material is lead- or lead-free and organic-inorganic or all-inorganic.

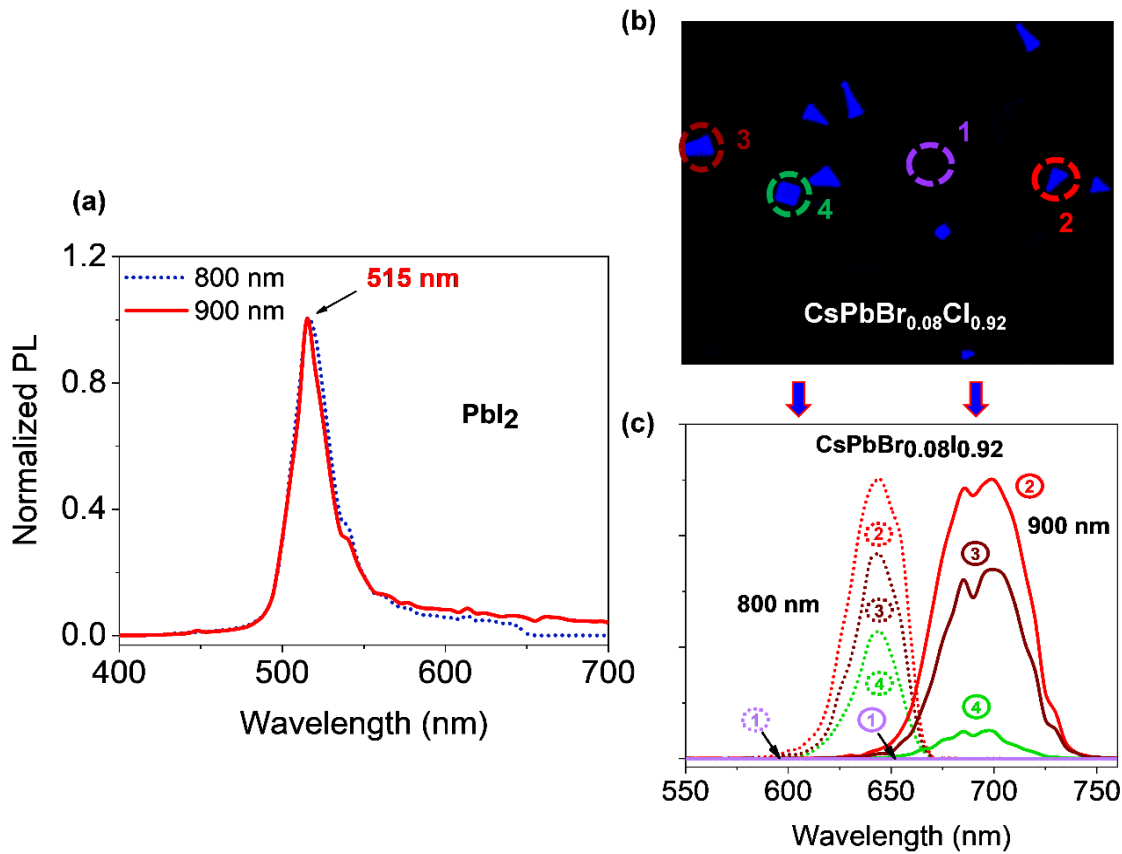

*Figure 3| Verifying the reversible PL phenomenon. a,* No exhibition of the reversible PL peak phenomenon in $PbI_2$ film. *b,* Photographs of different structures of microplatelets in $CsPbBr_{0.08}I_{0.92}$ iodide single crystal perovskites alloy. *c,* Excitation wavelength-dependent reversible PL peak with an energy window between two peaks is observed in all structures (triangular (2), rectangle-like (3), and square-like (4)) of microplatelets for the $CsPbBr_{0.08}I_{0.92}$ perovskite alloy, while position (1) is intentionally selected on a place where there is no microstructure.

Furthermore, PL study is conducted for several micro-platelet structures of all-inorganic single-crystal perovskite alloys with different shapes to observe whether the phenomenon is limited to certain structures. To do so, two-dimensional PL spectroscopy intensity mapping was utilized to determine and select the individual micro-platelet structure of $CsPbBr_{0.08}I_{0.92}$ perovskite alloy. Then, the photoluminescence spectroscopy analysis is performed on the particular single-crystal micro-platelet, while the TPA wavelength is tuned between 800 nm and 900 nm. Figure

3(b) displays the PL images of several structures with triangular (2), square-like (3), and rectangle-like (4) shaped micro-platelet, while 3(c) presents the corresponding PL spectra exhibiting the same reversible PL peaks behavior between 700 nm (1.77 eV) and 645 nm (1.92 eV) for the TPA excitation wavelengths of 900 nm and 800 nm, respectively. Similar spectral evolution can be observed for $CsPbBr_{0.15}I_{0.85}$ iodide single-crystal perovskite alloy, whereas the $CsPbBr_{0.4}Cl_{0.6}$ non-iodide perovskite alloy does not show the same (see Fig. S3.). Note that a bare space (1) without any microplatelet deposited shows no PL emission (Fig. 3(c) purple spectrum), confirming that the PL is, in fact, originating from the structure but not from the substrate.

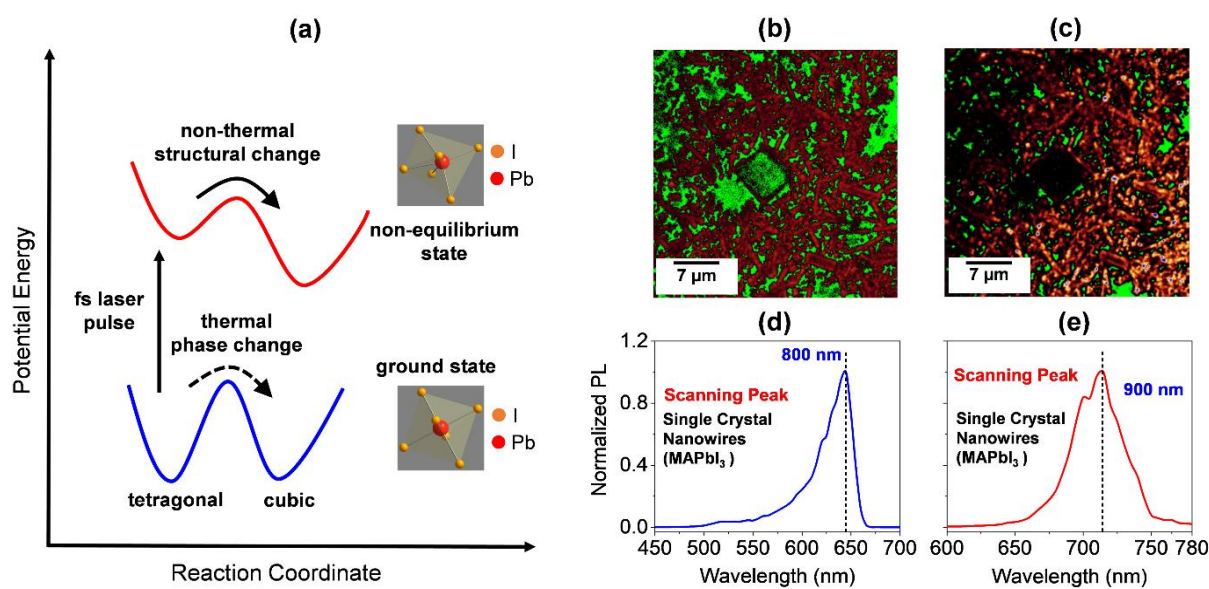

*Figure 4| Photoinduced structural change in perovskites. a, Schematic diagram illustrating the commonly studied thermal phase change (bottom) and photoinduced non-thermal structural change under the femtosecond laser excitation. The perovskite structure shown in the inset (bottom) can be structurally deformed as presented on the top inset figure due to the rotation of the iodine element around the lead [37]. (b, c), two-dimensional photoluminescence intensity mapping of $MAPbI_3$ single crystal nanowire under the TPA excitation laser beam of (b) 800 nm*

*and (c) 900 nm. **(d, e)**, the corresponding scanning PL peak at (d) 645 nm and (e)715 nm, respectively.*

We propose that the excitation wavelength-dependent reversible spectral evolution of PL peak and the reversible exhibition of low and high energy prominent peaks are attributed to the non-thermal photoinduced interim structural change in the iodide perovskites, in which the laser-induced rotation of iodine element plays an important role of deforming the perovskite structure and modulating the corresponding optical properties. In general, perovskites can exhibit several phases with different crystal structures and optoelectronic properties at different temperature [43–45]. For example, a tetragonal perovskite phase structure can be changed to a cubic structure along with the reaction coordinate by employing temperature effect as shown in Fig. 4(a) (bottom, left to right), where the potential energy surface is assumed to be the same [46]. However, as previously mentioned, fs laser light pulse can also cause a huge deformation in the bonding of iodine and lead atoms in perovskite within only ten trillionths of a second [37], while both the reaction coordinate and potential energy surface can be changed [46]. Moreover, A. Kumari et al. [47] and Wang et al. [40] reported that different emission energy could result from perovskites with different bond distances and connectivity in the network of infinitely extended lead-halide octahedra structures. At first, to confirm whether the energy excitation wavelength-dependent such phenomenon is articulated only from the lead-iodide but not being an inherent optical property of perovskite, lead-iodide ($PbI_2$) film was measured by the tunable fs laser source as depicted in Fig. 3(a). Only an emission peak at 515 nm is observed with no excitation wavelength dependent reversible PL or pronouncing multiple PL peaks for the higher energy laser excitation in $PbI_2$ film. Similar behavior was also measured in tin-iodide ($SnI_2$) results provided in the supplementary Fig. S4. These results reveal the possible contribution of iodine element in an infinitely extended Pb-I-Pb/Sn-I-Sn framework only, when it comes along with the perovskites. This is because that the charge anisotropy of organic/inorganic cation

(organic: $CH_3NH_3^+$ ($MA^+$) or $NH_2CH_3NH_2^+$($FA^+$), inorganic: Cs) facilitates the rotation of iodine within the cuboctahedral lattice cavities of anharmonic double-well (see Fig. 4(a)) halide-perovskite [48].

In the next step, a two-dimensional (2D) photoluminescence intensity mapping is performed to directly observe the image of the illuminated samples while tuning the fs laser wavelengths between 750 nm and 900 nm. During the experiment, a focused laser beam continuously scans over the sample, where the spot size of the laser beam and the scanning area was ~1-2 μm and >400 nm, respectively, resulting in scanning images and emission intensity profile in grayscale. The spectra acquisition, emission intensity evolution, and all other subsequent analysis are based on the obtained false-color images. It should be noted here that high precision is required in focusing the sample and adjusting the optimal laser intensity simultaneously to monitor such images, especially for the polycrystalline film type perovskites prepared by solution-processed method (see Fig.1 (c,d)), since the $MAPbI_3$ samples are not stable at high laser light illumination. Interestingly, we have observed distinctly different microstructures for $MAPbI_3$ polycrystalline film [Fig. 1 (c, d)], $MAPbI_3$ single-crystal nanowire [Fig. 4 (b, c)], and $CsPbBr_{0.41}I_{0.59}$ single-crystal alloy [Fig. S5 (a, b)], under TPA of fs laser excitations of 800 nm and 900 nm. The mechanism needs to be further investigated why the intensity mapping photoluminescence images illuminated by two different TPA wavelengths are different, and whether such changes have any influence on resolving reversible PL. At this moment, we believe that it should be related to the laser-induced structural distortion. Note that the fs laser-induced PL phenomenon is caused by non-equilibrium phase transition under intense ultra-fast laser illumination and the abrupt jump requires an energy sufficient to overcome an activation barrier to go from one metastable state for exhibiting high energy PL peaks under higher excitation energy.

In summary, a phenomenon of reversible PL peak was experimentally observed in iodide-based perovskites, while the samples were excited by a femtosecond laser with TPA energy higher than a threshold energy. This photoinduced energy may be enough to initiate the rotation of the iodine element around the Pb/Sn, resulting in the distortion in the perovskite structures. Moreover, a relatively high energy peak was also exhibited under a relatively shorter/higher excitation wavelength/energy. Interestingly, this process was also observed as reversible with the excitation wavelength. In addition, this intrinsic property can be realized in all the perovskites regardless of organic-inorganic and all-inorganic lead- or lead-free, but only for perovskite with iodine as one of the halide elements. Moreover, the synthesis method and the microstructure of the perovskites do not affect such reversible PL peak. We believe that our finding of non-thermal and non-equilibrium ultrafast phase transition in all kinds of iodide perovskites coupled with femtosecond two-photon absorption process or multiple terahertz light-pulse sequences will provide relevant fundamental insights for perovskite materials based ultrafast optical data processing and for the next generation of ultra-high-speed phase-change random access memory (PCRAM). In addition, this tunable and reversible optoelectronic properties of perovskites might facilitate research related to tunable laser and light emitting diodes, fast transport of carriers, interface engineering, and degradation measurement of solar cells for improving the energy conversion efficiency.

**Data availability**

The data that support the plots within this paper and other finding of this study are available from the corresponding authors upon reasonable request.

**Acknowledgments**

This work was financially supported by the grant from Shenzhen Municipal Science and Technology projects (Grant No. JCYJ201605313001154), The Hong Kong Research Grants

Council of Hong Kong (GRF 152109/16E) and The Hong Kong Polytechnic University (Project number: G-YBVG, 1-ZVGH). MKH and KMY acknowledge the financial support by a CityU SGP Grant (No. 9380076). HS acknowledges the financial support by the grant from Shenzhen Municipal Science and Technology Projects No. JCYJ20160531192714636. The access to the Leica TCS SP8 MP system was provided by the University Research Facility in Life Sciences (ULS) of The Hong Kong Polytechnic University. Thanks to the technical persons in ULS for their technical support as well.

**Supporting Information**

Supporting Information is available from the author.

**References**

[1]   G. Xing, N. Mathews, S.S. Lim, N. Yantara, X. Liu, D. Sabba, M. Grätzel, S. Mhaisalkar, T.C. Sum, Low-temperature solution-processed wavelength-tunable perovskites for lasing, Nat. Mater. 13 (2014) 476–480. doi:10.1038/nmat3911.

[2]   H. Zhu, Y. Fu, F. Meng, X. Wu, Z. Gong, Q. Ding, M. V. Gustafsson, M.T. Trinh, S. Jin, X.-Y. Zhu, Lead halide perovskite nanowire lasers with low lasing thresholds and high quality factors, Nat. Mater. 14 (2015) 636–642. doi:10.1038/nmat4271.

[3]   S.D. Stranks, H.J. Snaith, Metal-halide perovskites for photovoltaic and light-emitting devices, Nat. Nanotechnol. 10 (2015) 391–402. doi:10.1038/nnano.2015.90.

[4]   N. Arora, M.I. Dar, M. Hezam, W. Tress, G. Jacopin, T. Moehl, P. Gao, A.S. Aldwayyan, B. Deveaud, M. Grätzel, M.K. Nazeeruddin, Photovoltaic and Amplified Spontaneous Emission Studies of High-Quality Formamidinium Lead Bromide Perovskite Films, Adv. Funct. Mater. 26 (2016) 2846–2854. doi:10.1002/adfm.201504977.

[5]   K. Wang, W. Sun, J. Li, Z. Gu, S. Xiao, Q. Song, Unidirectional Lasing Emissions

from $CH_3NH_3PbBr_3$ Perovskite Microdisks, ACS Photonics. 3 (2016) 1125–1130. doi:10.1021/acsphotonics.6b00209.

[6]  Z. Duan, S. Wang, N. Yi, Z. Gu, Y. Gao, Q. Song, S. Xiao, Miscellaneous Lasing Actions in Organo-Lead Halide Perovskite Films, ACS Appl. Mater. Interfaces. 9 (2017) 20711–20718. doi:10.1021/acsami.7b01383.

[7]  C. Barugkin, J. Cong, T. Duong, S. Rahman, H.T. Nguyen, D. Macdonald, T.P. White, K.R. Catchpole, Ultralow Absorption Coefficient and Temperature Dependence of Radiative Recombination of $CH_3NH_3PbI_3$ Perovskite from Photoluminescence, J. Phys. Chem. Lett. 6 (2015) 767–772. doi:10.1021/acs.jpclett.5b00044.

[8]  X. Wang, M. Shoaib, X. Wang, X. Zhang, M. He, Z. Luo, W. Zheng, H. Li, T. Yang, X. Zhu, L. Ma, A. Pan, High-Quality In-Plane Aligned $CsPbX_3$ Perovskite Nanowire Lasers with Composition-Dependent Strong Exciton–Photon Coupling, ACS Nano. 12 (2018) 6170–6178. doi:10.1021/acsnano.8b02793.

[9]  P. Liu, X. He, J. Ren, Q. Liao, J. Yao, H. Fu, Organic–Inorganic Hybrid Perovskite Nanowire Laser Arrays, ACS Nano. 11 (2017) 5766–5773. doi:10.1021/acsnano.7b01351.

[10] J. Xing, X.F. Liu, Q. Zhang, S.T. Ha, Y.W. Yuan, C. Shen, T.C. Sum, Q. Xiong, Vapor Phase Synthesis of Organometal Halide Perovskite Nanowires for Tunable Room-Temperature Nanolasers, Nano Lett. 15 (2015) 4571–4577. doi:10.1021/acs.nanolett.5b01166.

[11] B. Yang, X. Mao, S. Yang, Y. Li, Y. Wang, M. Wang, W.-Q. Deng, K.-L. Han, Low Threshold Two-Photon-Pumped Amplified Spontaneous Emission in $CH_3NH_3PbBr_3$ Microdisks, ACS Appl. Mater. Interfaces. 8 (2016) 19587–19592. doi:10.1021/acsami.6b04246.

[12] J.A. Sichert, Y. Tong, N. Mutz, M. Vollmer, S. Fischer, K.Z. Milowska, R. García Cortadella, B. Nickel, C. Cardenas-Daw, J.K. Stolarczyk, A.S. Urban, J. Feldmann,

Quantum Size Effect in Organometal Halide Perovskite Nanoplatelets, Nano Lett. 15 (2015) 6521–6527. doi:10.1021/acs.nanolett.5b02985.

[13] P. Tyagi, S.M. Arveson, W.A. Tisdale, Colloidal Organohalide Perovskite Nanoplatelets Exhibiting Quantum Confinement, J. Phys. Chem. Lett. 6 (2015) 1911–1916. doi:10.1021/acs.jpclett.5b00664.

[14] J. Chen, K. Žídek, P. Chábera, D. Liu, P. Cheng, L. Nuuttila, M.J. Al-Marri, H. Lehtivuori, M.E. Messing, K. Han, K. Zheng, T. Pullerits, Size- and Wavelength-Dependent Two-Photon Absorption Cross-Section of CsPbBr 3 Perovskite Quantum Dots, J. Phys. Chem. Lett. 8 (2017) 2316–2321. doi:10.1021/acs.jpclett.7b00613.

[15] Q. Han, W. Wu, W. Liu, Y. Yang, The peak shift and evolution of upconversion luminescence from CsPbBr 3 nanocrystals under femtosecond laser excitation, RSC Adv. 7 (2017) 35757–35764. doi:10.1039/C7RA06211G.

[16] L. Wang, G.D. Yuan, R.F. Duan, F. Huang, T.B. Wei, Z.Q. Liu, J.X. Wang, J.M. Li, Tunable bandgap in hybrid perovskite CH 3 NH 3 Pb(Br 3−y X y ) single crystals and photodetector applications, AIP Adv. 6 (2016) 045115. doi:10.1063/1.4948312.

[17] I.E. Castelli, J.M. García-Lastra, K.S. Thygesen, K.W. Jacobsen, Bandgap calculations and trends of organometal halide perovskites, APL Mater. 2 (2014) 081514. doi:10.1063/1.4893495.

[18] G. Xing, N. Mathews, S. Sun, S.S. Lim, Y.M. Lam, M. Gratzel, S. Mhaisalkar, T.C. Sum, Long-Range Balanced Electron- and Hole-Transport Lengths in Organic-Inorganic CH3NH3PbI3, Science (80-. ). 342 (2013) 344–347. doi:10.1126/science.1243167.

[19] Y. Li, W. Yan, Y. Li, S. Wang, W. Wang, Z. Bian, L. Xiao, Q. Gong, Direct Observation of Long Electron-Hole Diffusion Distance in CH3NH3PbI3 Perovskite Thin Film, Sci. Rep. 5 (2015) 14485. doi:10.1038/srep14485.

[20] D. Shi, V. Adinolfi, R. Comin, M. Yuan, E. Alarousu, A. Buin, Y. Chen, S. Hoogland,

A. Rothenberger, K. Katsiev, Y. Losovyj, X. Zhang, P.A. Dowben, O.F. Mohammed, E.H. Sargent, O.M. Bakr, Low trap-state density and long carrier diffusion in organolead trihalide perovskite single crystals, Science (80-. ). 347 (2015) 519–522. doi:10.1126/science.aaa2725.

[21] M.A. Green, Y. Jiang, A.M. Soufiani, A. Ho-Baillie, Optical Properties of Photovoltaic Organic–Inorganic Lead Halide Perovskites, J. Phys. Chem. Lett. 6 (2015) 4774–4785. doi:10.1021/acs.jpclett.5b01865.

[22] Y. Jiang, M.A. Green, R. Sheng, A. Ho-Baillie, Room temperature optical properties of organic–inorganic lead halide perovskites, Sol. Energy Mater. Sol. Cells. 137 (2015) 253–257. doi:10.1016/j.solmat.2015.02.017.

[23] Y. Bekenstein, B.A. Koscher, S.W. Eaton, P. Yang, A.P. Alivisatos, Highly Luminescent Colloidal Nanoplates of Perovskite Cesium Lead Halide and Their Oriented Assemblies, J. Am. Chem. Soc. 137 (2015) 16008–16011. doi:10.1021/jacs.5b11199.

[24] L. Protesescu, S. Yakunin, M.I. Bodnarchuk, F. Krieg, R. Caputo, C.H. Hendon, R.X. Yang, A. Walsh, M. V. Kovalenko, Nanocrystals of Cesium Lead Halide Perovskites ($CsPbX_3$, X = Cl, Br, and I): Novel Optoelectronic Materials Showing Bright Emission with Wide Color Gamut, Nano Lett. 15 (2015) 3692–3696. doi:10.1021/nl5048779.

[25] A. Swarnkar, R. Chulliyil, V.K. Ravi, M. Irfanullah, A. Chowdhury, A. Nag, Colloidal $CsPbBr_3$ Perovskite Nanocrystals: Luminescence beyond Traditional Quantum Dots, Angew. Chemie Int. Ed. 54 (2015) 15424–15428. doi:10.1002/anie.201508276.

[26] F. Deschler, M. Price, S. Pathak, L.E. Klintberg, D.-D. Jarausch, R. Higler, S. Hüttner, T. Leijtens, S.D. Stranks, H.J. Snaith, M. Atatüre, R.T. Phillips, R.H. Friend, High Photoluminescence Efficiency and Optically Pumped Lasing in Solution-Processed Mixed Halide Perovskite Semiconductors, J. Phys. Chem. Lett. 5 (2014) 1421–1426.

doi:10.1021/jz5005285.

[27] C.S. Ponseca, T.J. Savenije, M. Abdellah, K. Zheng, A. Yartsev, T. Pascher, T. Harlang, P. Chabera, T. Pullerits, A. Stepanov, J.-P. Wolf, V. Sundström, Organometal Halide Perovskite Solar Cell Materials Rationalized: Ultrafast Charge Generation, High and Microsecond-Long Balanced Mobilities, and Slow Recombination, J. Am. Chem. Soc. 136 (2014) 5189–5192. doi:10.1021/ja412583t.

[28] A. Kojima, K. Teshima, Y. Shirai, T. Miyasaka, Organometal Halide Perovskites as Visible-Light Sensitizers for Photovoltaic Cells, J. Am. Chem. Soc. 131 (2009) 6050–6051. doi:10.1021/ja809598r.

[29] M.A. Green, Y. Hishikawa, E.D. Dunlop, D.H. Levi, J. Hohl-Ebinger, A.W.Y. Ho-Baillie, Solar cell efficiency tables (version 51), Prog. Photovoltaics Res. Appl. 26 (2018) 3–12. doi:10.1002/pip.2978.

[30] N.-G. Park, Perovskite solar cells: an emerging photovoltaic technology, Mater. Today. 18 (2015) 65–72. doi:10.1016/j.mattod.2014.07.007.

[31] K.A. Bush, A.F. Palmstrom, Z.J. Yu, M. Boccard, R. Cheacharoen, J.P. Mailoa, D.P. McMeekin, R.L.Z. Hoye, C.D. Bailie, T. Leijtens, I.M. Peters, M.C. Minichetti, N. Rolston, R. Prasanna, S. Sofia, D. Harwood, W. Ma, F. Moghadam, H.J. Snaith, T. Buonassisi, Z.C. Holman, S.F. Bent, M.D. McGehee, 23.6%-efficient monolithic perovskite/silicon tandem solar cells with improved stability, Nat. Energy. 2 (2017) 17009. doi:10.1038/nenergy.2017.9.

[32] F. Sahli, J. Werner, B.A. Kamino, M. Bräuninger, R. Monnard, B. Paviet-Salomon, L. Barraud, L. Ding, J.J. Diaz Leon, D. Sacchetto, G. Cattaneo, M. Despeisse, M. Boccard, S. Nicolay, Q. Jeangros, B. Niesen, C. Ballif, Fully textured monolithic perovskite/silicon tandem solar cells with 25.2% power conversion efficiency, Nat. Mater. 17 (2018) 820–826. doi:10.1038/s41563-018-0115-4.

[33] M.I. Hossain, W. Qarony, V. Jovanov, Y.H. Tsang, D. Knipp, Nanophotonic design of

perovskite/silicon tandem solar cells, J. Mater. Chem. A. 6 (2018) 3625–3633. doi:10.1039/C8TA00628H.

[34]  S. Wang, Y. Jiang, E.J. Juarez-Perez, L.K. Ono, Y. Qi, Accelerated degradation of methylammonium lead iodide perovskites induced by exposure to iodine vapour, Nat. Energy. 2 (2017) 16195. doi:10.1038/nenergy.2016.195.

[35]  F. Liu, Q. Dong, M.K. Wong, A.B. Djurišić, A. Ng, Z. Ren, Q. Shen, C. Surya, W.K. Chan, J. Wang, A.M.C. Ng, C. Liao, H. Li, K. Shih, C. Wei, H. Su, J. Dai, Is Excess PbI 2 Beneficial for Perovskite Solar Cell Performance?, Adv. Energy Mater. 6 (2016) 1502206. doi:10.1002/aenm.201502206.

[36]  G. Niu, X. Guo, L. Wang, Review of recent progress in chemical stability of perovskite solar cells, J. Mater. Chem. A. 3 (2015) 8970–8980. doi:10.1039/C4TA04994B.

[37]  X. Wu, L.Z. Tan, X. Shen, T. Hu, K. Miyata, M.T. Trinh, R. Li, R. Coffee, S. Liu, D.A. Egger, I. Makasyuk, Q. Zheng, A. Fry, J.S. Robinson, M.D. Smith, B. Guzelturk, H.I. Karunadasa, X. Wang, X. Zhu, L. Kronik, A.M. Rappe, A.M. Lindenberg, Light-induced picosecond rotational disordering of the inorganic sublattice in hybrid perovskites, Sci. Adv. 3 (2017) e1602388. doi:10.1126/sciadv.1602388.

[38]  K. Leng, I. Abdelwahab, I. Verzhbitskiy, M. Telychko, L. Chu, W. Fu, X. Chi, N. Guo, Z. Chen, Z. Chen, C. Zhang, Q.-H. Xu, J. Lu, M. Chhowalla, G. Eda, K.P. Loh, Molecularly thin two-dimensional hybrid perovskites with tunable optoelectronic properties due to reversible surface relaxation, Nat. Mater. 17 (2018) 908–914. doi:10.1038/s41563-018-0164-8.

[39]  M. Hase, P. Fons, K. Mitrofanov, A. V. Kolobov, J. Tominaga, Femtosecond structural transformation of phase-change materials far from equilibrium monitored by coherent phonons, Nat. Commun. 6 (2015) 8367. doi:10.1038/ncomms9367.

[40]  Y. Wang, T. Gould, J.F. Dobson, H. Zhang, H. Yang, X. Yao, H. Zhao, Density functional theory analysis of structural and electronic properties of orthorhombic

perovskite CH 3 NH 3 PbI 3, Phys. Chem. Chem. Phys. 16 (2014) 1424–1429. doi:10.1039/C3CP54479F.

[41] J. Li, S. Zhang, H. Dong, X. Yuan, X. Jiang, J. Wang, L. Zhang, Two-photon absorption and emission in CsPb(Br/I) 3 cesium lead halide perovskite quantum dots, CrystEngComm. 18 (2016) 7945–7949. doi:10.1039/C6CE01864E.

[42] T.M. Brenner, D.A. Egger, L. Kronik, G. Hodes, D. Cahen, Hybrid organic—inorganic perovskites: low-cost semiconductors with intriguing charge-transport properties, Nat. Rev. Mater. 1 (2016) 15007. doi:10.1038/natrevmats.2015.7.

[43] P.S. Whitfield, N. Herron, W.E. Guise, K. Page, Y.Q. Cheng, I. Milas, M.K. Crawford, Structures, Phase Transitions and Tricritical Behavior of the Hybrid Perovskite Methyl Ammonium Lead Iodide, Sci. Rep. 6 (2016) 35685. doi:10.1038/srep35685.

[44] C. Chen, X. Hu, W. Lu, S. Chang, L. Shi, L. Li, H. Zhong, J.-B. Han, Elucidating the phase transitions and temperature-dependent photoluminescence of MAPbBr 3 single crystal, J. Phys. D. Appl. Phys. 51 (2018) 045105. doi:10.1088/1361-6463/aaa0ed.

[45] T.J. Jacobsson, L.J. Schwan, M. Ottosson, A. Hagfeldt, T. Edvinsson, Determination of Thermal Expansion Coefficients and Locating the Temperature-Induced Phase Transition in Methylammonium Lead Perovskites Using X-ray Diffraction, Inorg. Chem. 54 (2015) 10678–10685. doi:10.1021/acs.inorgchem.5b01481.

[46] K.H. Bennemann, Photoinduced phase transitions, J. Phys. Condens. Matter. 23 (2011) 073202. doi:10.1088/0953-8984/23/7/073202.

[47] A. Kumari, I. Singh, N. Prasad, S.K. Dixit, P.K. Rao, P.K. Bhatnagar, P.C. Mathur, C.S. Bhatia, S. Nagpal, Improving the efficiency of a poly(3-hexylthiophene)-CuInS2 photovoltaic device by incorporating graphene nanopowder, J. Nanophotonics. 8 (2014) 083092. doi:10.1117/1.JNP.8.083092.

[48] C. Katan, A.D. Mohite, J. Even, Entropy in halide perovskites, Nat. Mater. 17 (2018) 377–379. doi:10.1038/s41563-018-0070-0.

Supplementary

# Excitation Wavelength Dependent Reversible Photoluminescence Peak in Iodide Perovskites

Wayesh Qarony[1], Mohammad K. Hossain[3], Mohammad I. Hossain[1], Sainan Ma[1], Longhui Zeng[1], Kin Man Yu[3], Dietmar Knipp[4], Alberto Salleo[4], Huarui Sun[2], Cho Tung Yip[*, 2], Yuen Hong Tsang[*, 1]

1.) Department of Applied Physics, The Hong Kong Polytechnic University, Hung Hom, Kowloon, Hong Kong

2.) School of Science, Harbin Institute of Technology, Shenzhen Graduate School, Shenzhen, Guangdong 58055, China

3.) Department of Physics, City University of Hong Kong, Kowloon, Hong Kong

4.) Geballe Laboratory for Advanced Materials, Department of Materials Science and Engineering, Stanford University, Stanford, CA 94305, USA

*Corresponding authors: E-mail: yuen.tsang@polyu.edu.hk (Y. H. Tsang), h0260416@hit.edu.cn (C.T.Yip)

**This file includes:**

Figs. S1, S2, S3, S4, S5

Materials and Methods

Supplementary Text

References

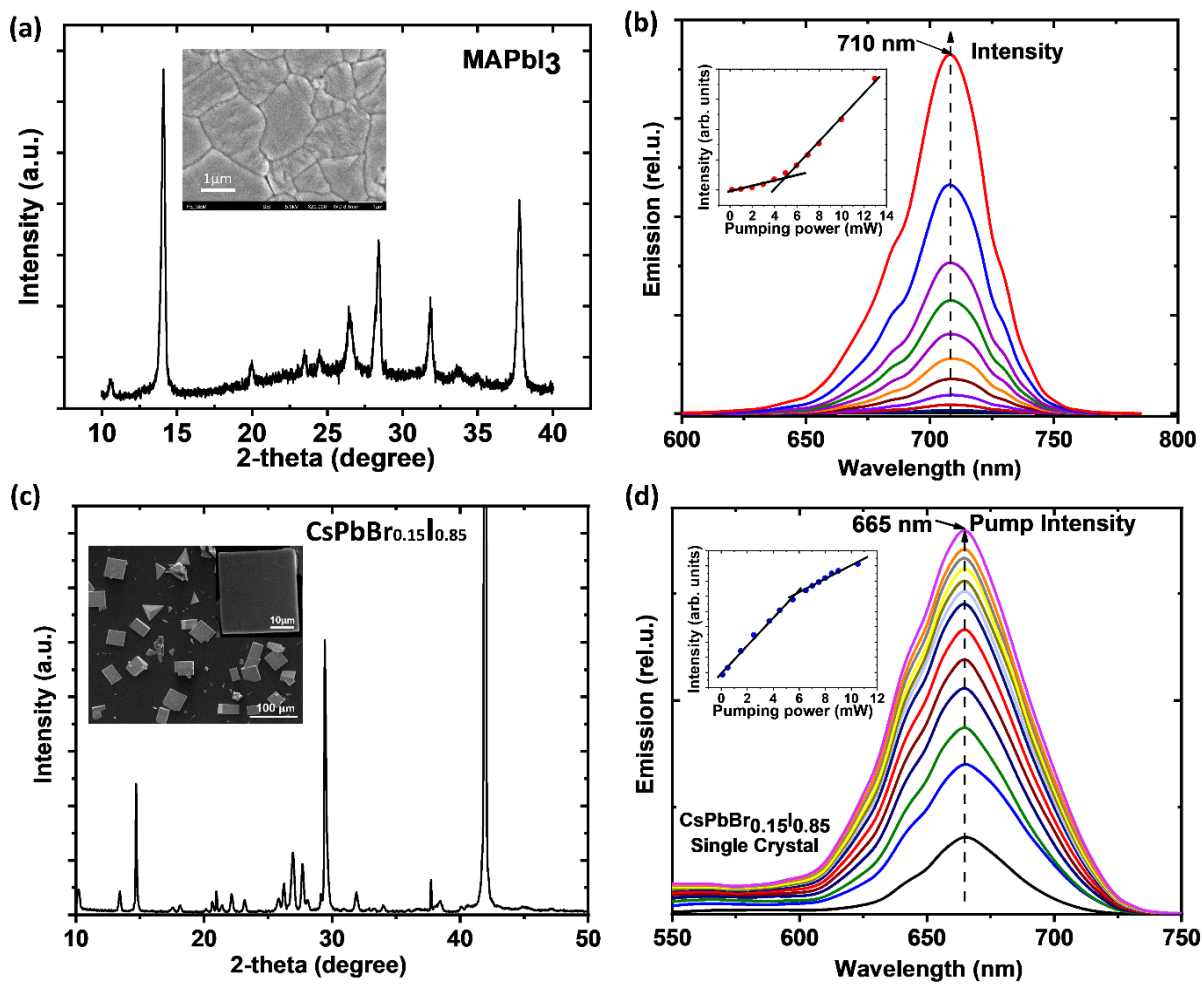

*Figure S1| Materials and lasing properties of the organic-inorganic and all-inorganic iodide perovskites (film and single crystal). a, XRD spectrum and SEM (inset) of the $CH_3NH_3PbI_3$ film prepared by two-step solution-processed method. b, Room temperature power-dependent PL spectra of $CH_3NH_3PbI_3$ film at 900 nm of TPA lasing wavelength. Inset is integrated output emission peak over the whole spectra range as a function of pumping power. The experiment data (red points) is well fitted by distinguished functions (black line). c, XRD spectrum and SEM (inset) of the $CsPbBr_{0.15}I_{0.85}$ microplatelets single crystal prepared by two-step chemical vapor deposition method. d, Room temperature power-dependent PL spectra of $CsPbBr_{0.15}I_{0.85}$ microplatelets single crystal at 900 nm of TPA lasing wavelength. Inset is integrated output emission over the whole spectra range as a function of pumping power. The experiment data (blue points) is well fitted by distinguished functions (black line).*

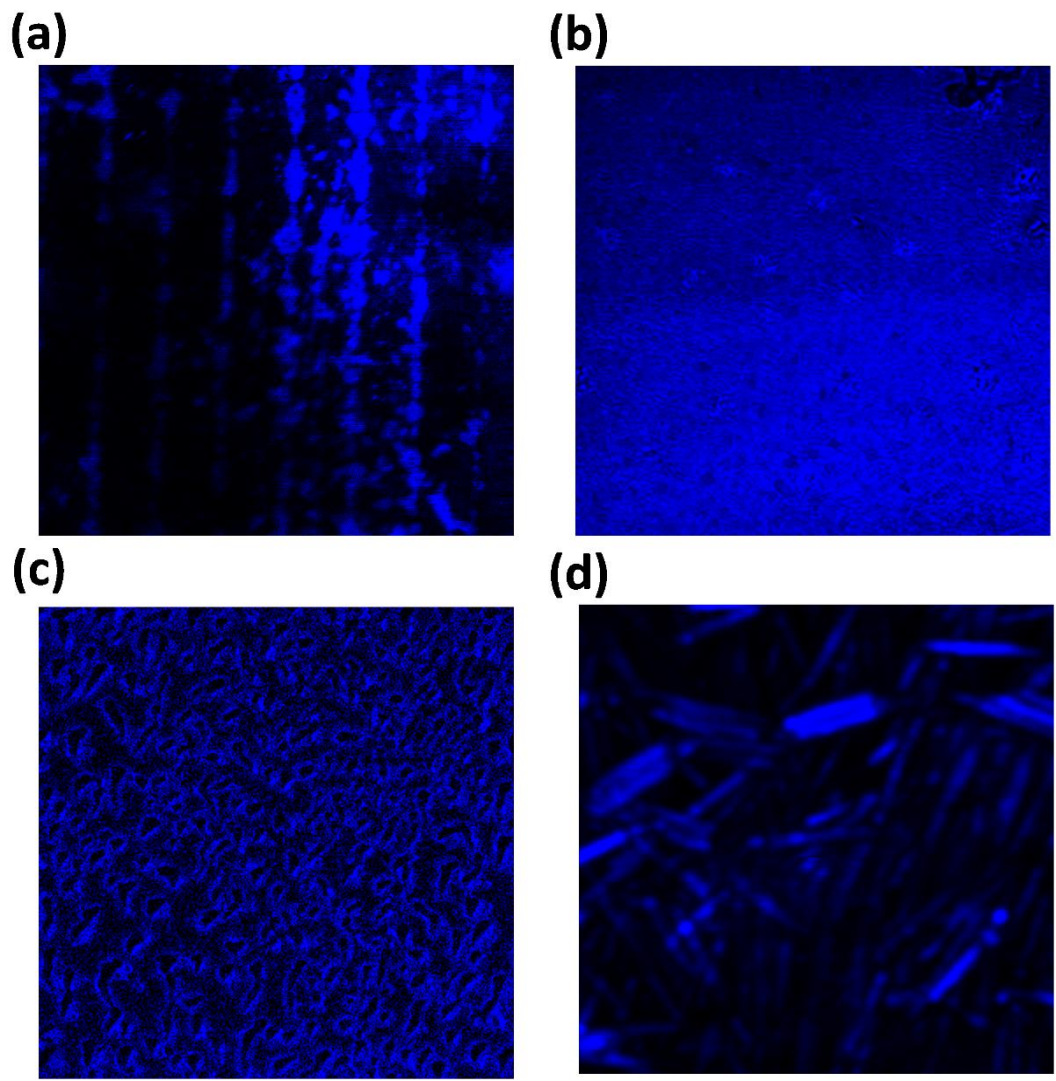

*Figure S2| Femtosecond laser illuminated samples. Photographs of (a) CsPbBr$_3$ film, (b) FA$_x$MA$_{1-x}$PbI$_3$ film, (c) FASnI$_3$, and (d) MAPbI$_3$ single crystal nanowire illuminated by a 900 nm laser beam.*

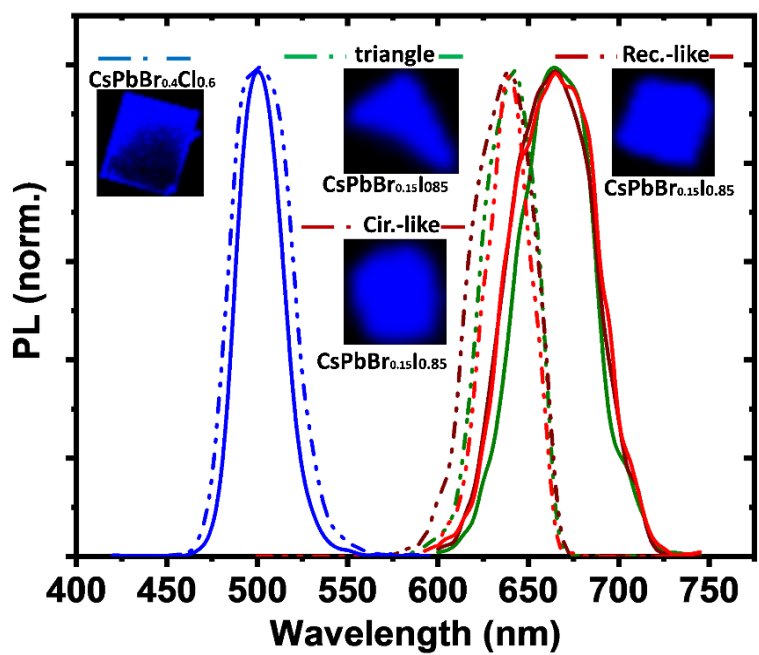

*Figure S3| Verifying the reversible PL peak phenomenon in single-crystal microplatelets of CsPbBr$_{0.15}$I$_{0.85}$ and CsPbBr$_{0.4}$Cl$_{0.6}$ perovskite alloy.* No exhibition of reversible PL peak in CsPbBr$_{0.4}$Cl$_{0.6}$ perovskite alloy. Reversible PL peak phenomenon is observed in all structures (triangle, rectangle-like, and circle-like) of microplatelets for the CsPbBr$_{0.15}$I$_{0.85}$ perovskite. A 680 nm of short-pass filter is used here to study the high energy peak rigorously.

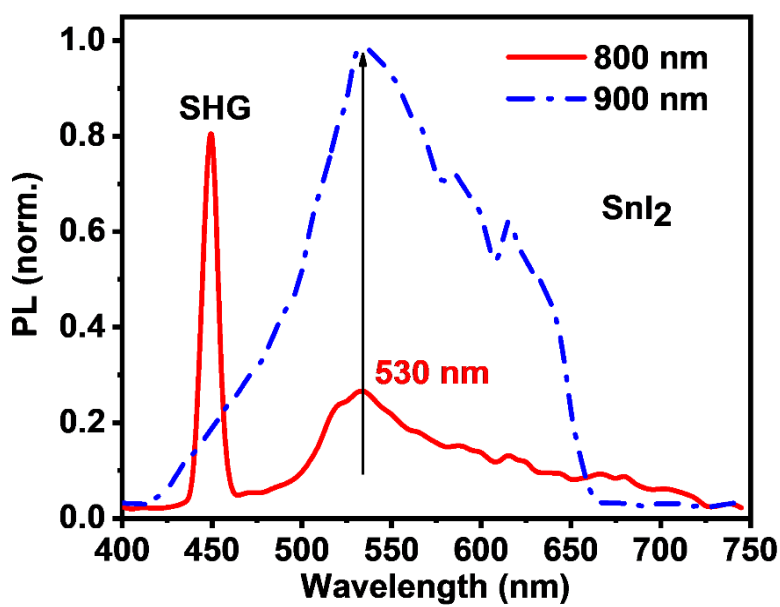

***Figure S4| Verifying the reversible PL peak phenomenon in SnI$_2$***, *No exhibition of reversible PL peak in only SnI$_2$ film. The peak exhibited at 450 nm for the excitation of 900 nm is a second harmonic generation (SHG).*

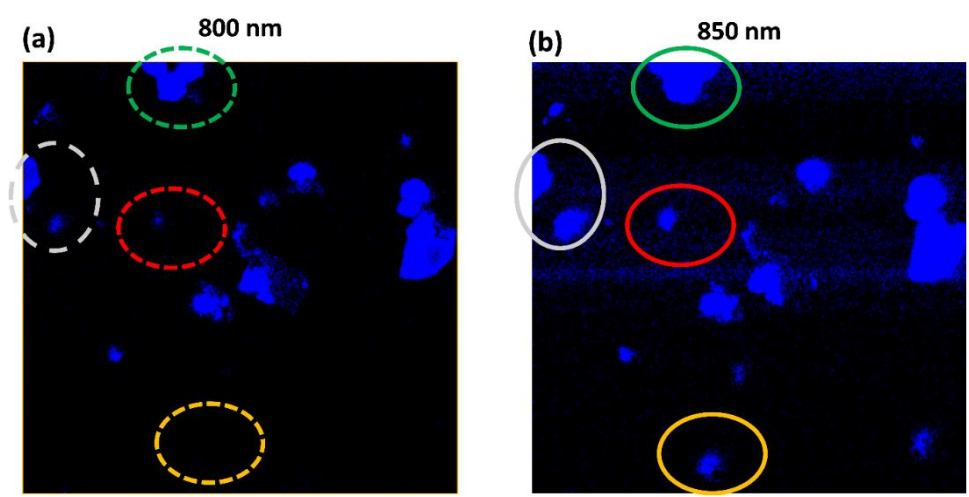

***Figure S5| Photoinduced structural change in perovskites.*** *two-dimensional photoluminescence intensity mapping of CsPbBr$_{0.41}$I$_{0.59}$ single crystal perovskite alloy excited by **(a)** an 800 nm TPA and **(b)** 850 nm TPA laser beam. The two images represent distinctly different in PL mapping structures, even though the intensity and the laser power were set to be the same.*

**Materials and Methods:**

Unless otherwise stated, all materials were purchased from Sigma-Aldrich or Alfa Aesar and used as received.

**Perovskites thin-film growth:**

**MAPbI$_3$ Film:**

The CH$_3$NH$_3$PbI$_3$ perovskite film is prepared using two-step method [S1-S2], where PbI$_2$ and CH$_3$NH$_3$I (MAI) are used as precursors. At first, lead iodide (PbI$_2$) film was deposited on 2.2 mm of FTO glass (or Silicon Oxide) substrate by spin coating (4000 r. p. m. for 30s) using PbI$_2$ (462 mg/1 mL) in DMF (anhydrous >99.98%, Sigma Aldrich) solution which was kept at 70$^0$ C for about 3 hours of vigorous stirring. Immediately after the PbI$_2$ deposition, MAI solution in 2-propanol (30 mg/1ml) (anhydrous >99.99%, Sigma Aldrich) was deposited by spin coating (2000 r. p. m.) and kept on the hot plate for about 6 hours of annealing. All the procedures including preparing solution are done inside glove box.

**FA$_x$MA$_{1-x}$PbI$_3$ film:**

The NH$_2$CH=NH$_2$PbI$_3$ perovskite film is prepared using two-step method [S1-S2], where PbI$_2$, MAI, and NH$_2$CH=NH$_2$I (FAI) are used as precursors. At first, a 150 mg of PbI$_2$ is dissolved with 220 ul of DMF (anhydrous >99.98%, Sigma Aldrich) and 20 ul of DMSO (anhydrous >99.98%, Sigma Aldrich). Then the solution was deposited on silicon oxide glass substrate by spin coating of 3000 r. p. m. for 30s. Immediately after the deposition, mixed FAI and MAI (20 mg of MAI and 25 mg of FAI in 1 ml of IPA) solution is deposited by spin coating with 5000 r. p. m. for 30 s. Then the samples are annealed for 1 hour. All the procedures including preparing solution are done inside glove box.

**CsPbBr$_3$ film:**

All the powder sources of lead bromide (PbBr$_2$, 99.99%) and cesium bromide (CsBr, 99.99%) were purchased from Sigma Aldrich and used as received without any further purification. At first, the solution of 1.0M lead bromide in DMF was spin-coated on the silicon oxide substrate at 3000 rpm for 30s and annealed at 70$^0$ C for 30 min. The solution was kept on the thermal stirrer at 70$^0$ C for 30 min before spin-coating. It was done inside a N$_2$ glovebox. Then, the PbBr$_2$ substrate was immersed in the solution of 0.07M CsBr in methanol at 70$^0$ C for 10 min, rinsed with IPA, spin-dried at 2000 rpm for 30 s and annealed at 100$^0$ C for 10 min inside fume cupboard to form the CsPbBr$_3$ perovskite film.

**FASnI$_3$ film:**

All the powder sources of tin-iodide (SnI$_2$, 99.999%) and formamidinium iodide (FAI, 99.99%) were purchased from Sigma Aldrich and used as received without any further purification. The FASnI$_3$ perovskite film is prepared using one-step method, where SnI$_2$ and NH$_2$CH=NH$_2$I (FAI) are used as precursors. 372.0 mg of SnI$_2$ and 172 mg of FAI are dissolved in 0.8 ml and 0.2 ml of DMF (anhydrous >99.98%, Sigma Aldrich) and DMSO (anhydrous >99.98%, Sigma Aldrich) with continuous stirring at room temperature inside a nitrogen-filled glovebox for about 3 hours. The FASnI$_3$ perovskite solution is spin-coated on a glass substrate covered by FTO at 4000 r.p.m for 60 s. The samples were then annealed at 70 °C for 20 min in the nitrogen-filled glovebox.

**Perovskite Alloy:**

Sapphire (Al$_2$O$_3$) was used as the substrates to grow the halide alloy. Before growth, the sapphire was cut into small pieces of 2 x 1 cm size and cleaned ultrasonically by turns in acetone, ethanol and deionized water for 10 min each in an ultrasonic bath and a nitrogen drier was used to dry the substrates. The perovskite alloy microplatelets were grown by a vapor phase approach using a home-built CVD system consisting of a horizontal furnace equipped with a quartz tube which hence connected to a carrier gas inlet and a pump down system. A boron nitride boat loaded with mixed powders of PbBr$_2$ and CsBr was used as the source of precursor materials during the growth of CsPbBr$_3$ alloy platelets and PbI$_2$ and CsI played the same role during the CVD synthesis of the quaternary CsPbBr$_{3(1-x)}$I$_{3x}$ alloy platelets. Initially, the precursor loaded boat was placed at the center of the heating zone of the furnace while the substrate (Al$_2$O$_3$) was placed in the downstream side of the heating zone of the furnace. As soon as the system was pumped down, a flow of high purity argon gas (Ar – 99.999%) at 100 sccm was introduced into the system for around 20 min to purge the oxygen, moisture or any other residual of the air in the tube. Attaining a stable optimized pressure of 7.0 Torr, the furnace was heated to 550°C at a rate of 30°C/min for growing the CsPbBr$_3$ layer and the growth was lasted for 30 min. The furnace was then allowed to cool down naturally to the room temperature. To realize the desired CsPbBr$_{3(1-x)}$I$_{3x}$ alloy, CsPbI$_3$ was grown in an optimized growth environment on the sapphire substrate that contain the CsPbBr$_3$ microplatelets following the same procedure as mentioned above. Flow of argon, as a carrier gas, was maintained at a standard rate throughout the growth process. At the end of the growth process when the furnace temperature was cooled down to the room temperature, the sample was collected and used for necessary characterizations.

**Optical Characterization method**

All the optical characterization, including the photoluminescence (PL) spectroscopy and 2D PL intensity mapping and imaging were performed using a functionally extended commercial multiphoton (MP) confocal laser system (TCS SP8, Leica). This confocal MP laser scanning system is combined with a Ti:sapphire femtosecond laser (Mai Tai HP, Spectra-Physics) system, where the laser pulse has a repetition rate and time duration of 80MHz and 200 fs, respectively. A beam scanning resonator along with programming applications in computer are utilized to control the scanning of the laser beam at the focal plane. A dry objective with a high NA of 0.95 is used for focusing the laser beam before reaching to the sample plane. The focused laser beam continuously scans over the sample, resulting in scanning images and emission intensity profile in gray scale. The spectra acquisition, emission intensity evolution, and all other subsequent analysis are based on the obtained false color images. As the nonlinear optical emission and imaging are occurred in the very small focus volume and the background intensity must be intrinsically concealed, the size of the pinholes between two lenses must be carefully optimized in normal scanning confocal mode yet are set as entirely open.

[S1] J.-W. Lee, N.-G. Park, Two-step deposition method for high-efficiency perovskite solar cells, MRS Bull. **2015**, 40, 654e659.

[S2] J.-H. Im, H.-S. Kim, N.-G. Park, Morphology-photovoltaic property correlation in perovskite solar cells: one-step versus two-step deposition of CH3NH3PbI3, APL Mater. **2014**, 2, 081510.

[S3] B. Saparov and D. B. Mitzi, Organic-Inorganic Perovskites: Structural Versatility for Functional Materials Design, Chem. Rev., **2016**, 116, 4558–4596.

[S4] J. L. Knutson, J. D. Martin and D. B. Mitzi, Tuning the Band Gap in Hybrid Tin Iodide Perovskite Semiconductors Using Structural Templating, Inorg. Chem.,**2005**, 44, 4699–4705.

[S5] Z. Xu, D. B. Mitzi, C. D. Dimitrakopoulos and K. R. Maxcy, Structurally Tailored Organic−Inorganic Perovskites: Optical Properties and Solution-Processed Channel Materials for Thin-Film Transistors, Inorg. Chem., **2003**, 42, 2031–2039.

[S6] Z. Xiao, W. Meng, J. Wang, D. B. Mitzi and Y. Yan, Searching for promising new perovskite-based photovoltaic absorbers: the importance of electronic dimensionality, Mater. Horiz., **2017**, 4, 206.